\documentclass[12pt]{article} 
\usepackage{graphicx}
\usepackage{epsfig} 
\usepackage{latexsym} 
\usepackage{latexsym}
\usepackage{amssymb} 
\usepackage{amsmath} 
\usepackage{cite}
\newcommand{\labell}[1]{\label{#1}}
\newcommand{\reef}[1]{(\ref{#1})}

\DeclareSymbolFont{AMSb}{U}{msb}{m}{n}
\DeclareMathSymbol{\IN}{\mathbin}{AMSb}{"4E}
\DeclareMathSymbol{\IZ}{\mathbin}{AMSb}{"5A}
\DeclareMathSymbol{\IR}{\mathbin}{AMSb}{"52}
\DeclareMathSymbol{\Q}{\mathbin}{AMSb}{"51}
\DeclareMathSymbol{\II}{\mathbin}{AMSb}{"49}
\DeclareMathSymbol{\IC}{\mathbin}{AMSb}{"43}
\DeclareMathSymbol{\IP}{\mathbin}{AMSb}{"50}
\DeclareMathSymbol{\IH}{\mathbin}{AMSb}{"48}
\DeclareMathSymbol\IA{\mathalpha}{AMSb}{"41}
\DeclareMathSymbol\IS{\mathalpha}{AMSb}{"53}

\def\Q{{\cal Q}}

\begin{document}

\begin{flushright}
USC-04-07\\                                       
\end{flushright}
\bigskip
\begin{center} {\Large \bf Operators with Large Quantum Numbers,}

\bigskip\bigskip

{\Large \bf Spinning Strings, and  Giant Gravitons}

\end{center}

\bigskip
\bigskip
\bigskip

\centerline{\bf Veselin Filev, Clifford V.
  Johnson\footnote{Also: Visiting Professor
    at the Centre for Particle Theory, Department of Mathematical
    Sciences, University of Durham, Durham DH1 3LE, England.}}

\bigskip
\bigskip
\bigskip

  \centerline{\it Department of Physics and Astronomy }
\centerline{\it University of
Southern California}
\centerline{\it Los Angeles, CA 90089-0484, U.S.A.}

\medskip

\centerline{\small \tt filev@usc.edu, johnson1@usc.edu}
\bigskip
\bigskip


\bigskip
\bigskip


\begin{abstract}
  We study the behaviour of spinning strings in the background of
  various distributions of smeared giant gravitons in supergravity.
  This gives insights into the behaviour of operators of high
  dimension, spin and R--charge. Using a new coordinate system
  recently presented in the literature, we find that it is
  particularly natural to prepare backgrounds in which the probe
  operators develop a variety of interesting new behaviours.  Among
  these are the possession of orbital angular momentum as well as
  spin, the breakdown of logarithmic scaling of dimension with spin in
  the high spin regime, and novel splitting/fractionation processes.

\end{abstract}
\newpage \baselineskip=18pt \setcounter{footnote}{0}

\section{Introduction}
\label{sec:introduction}

\subsection{Background}

Recently, in ref.\cite{Lin:2004nb} a new coordinate system was
discovered which allows for a rather general description of the
type~IIB supergravity duals of 1/2--BPS states in ${\cal N}=4$
supersymmetric Yang--Mills theory. Specifically, metric is:
\begin{eqnarray}
ds^2&=&-h^{-2}(dt+V_idx^i)^2+h^2(dy^2+dx^i dx^i)+ye^Gd\Omega_3^2+ye^{-G}d{\tilde\Omega}_3^2\ ,\nonumber\\
h^{-2}&=&2y\cosh G\ ,\quad y\partial_y V_i=\epsilon_{ij}\partial_j z\ ,\nonumber\\
&&\hskip-2cm \quad y(\partial_iV_j-\partial_jV_i)=\epsilon_{ij}\partial_y z\ ,\quad z=\frac{1}{2}\tanh G\ ,
\labell{eq:bigmetric}
\end{eqnarray}
where $i,j=1,2$\ , and $z$ obeys the equation
\begin{equation}
\partial_i\partial_j z+y\partial_y\left(\frac{\partial_y z}{y}\right)=0\ .
\labell{eq:diffy}
\end{equation}
There is no axion or dilaton field, nor are the three--form potentials
non--zero. The five--form field strength is non--zero, but we will not
list it here.

Notably, the $(x^1,x^2)$ plane defined by setting the coordinate $y$
to zero is isomorphic to the phase space of the fermionic description
of 1/2--BPS states introduced in ref.\cite{Berenstein:2004kk}. On this
plane, the function $z$ must take one of the values $\pm\frac{1}{2}$
in order for the metric to be non--singular. Imagine colouring the
plane black where $z=-\frac12$ and white where $z=+\frac12$. Then one
has a patchwork of shapes on the plane giving the boundary condition
for the function $z$. The differential equation~\reef{eq:diffy}
supplies a solution for $z$ everywhere and this gives the entire ten
dimensional metric.

The plane is to be thought of as a realization of the fermionic phase
space of the matrix model description of the 1/2--BPS states as given
in ref.\cite{Berenstein:2004kk}. The simplest configuration is a
circular disc centered at the origin. This Fermi droplet, filled out
to radius $r_0$ (determined by the fact that there are $N$ fermions in
the droplet and that there is a minimum area allowed a single
fermion), is in fact simply AdS$_5\times S^5$ with radius $R^2=r_0$,
which is set by the $N$ of the ${\cal N}=4$ supersymmetric $SU(N)$
gauge theory to which it is dual in the usual way:
\begin{equation}
R^4=4\pi(\alpha^\prime)^2 g_s N=2(\alpha^\prime)^2\lambda\ ;\qquad \lambda=g^2_{\rm YM}N\ .
\end{equation}

\begin{figure}[ht]
\begin{center}
\includegraphics[scale=0.25]{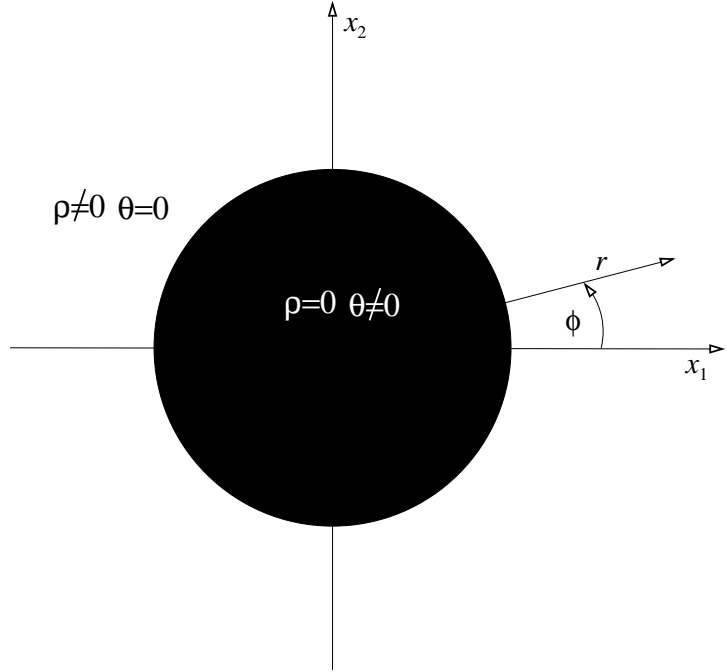}
\end{center}
\caption{\small The  configuration which corresponds to AdS$_5\times S^5$.}
\label{fig:disc}
\end{figure}

Defining cylindrical polars $(r,\phi)$ in the $(x_1,x_2)$ plane, the
coordinates relate to the familiar global AdS$_5\times S^5$ coordinates
in the following way: 
\begin{eqnarray} y=r_0\sinh\!\rho\sin\theta\
,\qquad r=r_0\cosh\!\rho\cos\theta \ ,\quad \phi={\tilde\phi}+t\ ,
\labell{eq:cov}
\end{eqnarray} 
where the metric is:
\begin{equation}
ds^2=r_0\left(d\rho^2-\cosh^2\rho dt^2+\sinh^2\rho
d\Omega_3^2+d\theta^2+\cos^2\theta d{\tilde\phi}^2+\sin^2\theta
d{\tilde\Omega}_3^2\right)\ .  
\end{equation}
Inside the disc,(see figure~\ref{fig:disc}), $z=-\frac12$, and
$\rho=0$, so $r$ runs from $0$ to $r_0$ as $\theta$ runs from
$\frac{\pi}{2}$ to~$0$. The $S^3$ of AdS$_5$ is collapsed in the
interior, while that of the $S^5$, denoted ${\tilde S}^3$, has radius
set by $\sqrt{r_0}\sin\theta$. So it shrinks to zero on the boundary
of the disc. So $\theta=0$ is the pole of the~$S^5$. Beyond the disc,
$\theta=0$ and $\rho$ runs from $0$ to $\infty$. The $S^3$ has radius
$\sqrt{r_0}\sinh\rho$, and we construct AdS$_5$.

\begin{figure}[htb]
\begin{center}
\includegraphics[scale=0.2]{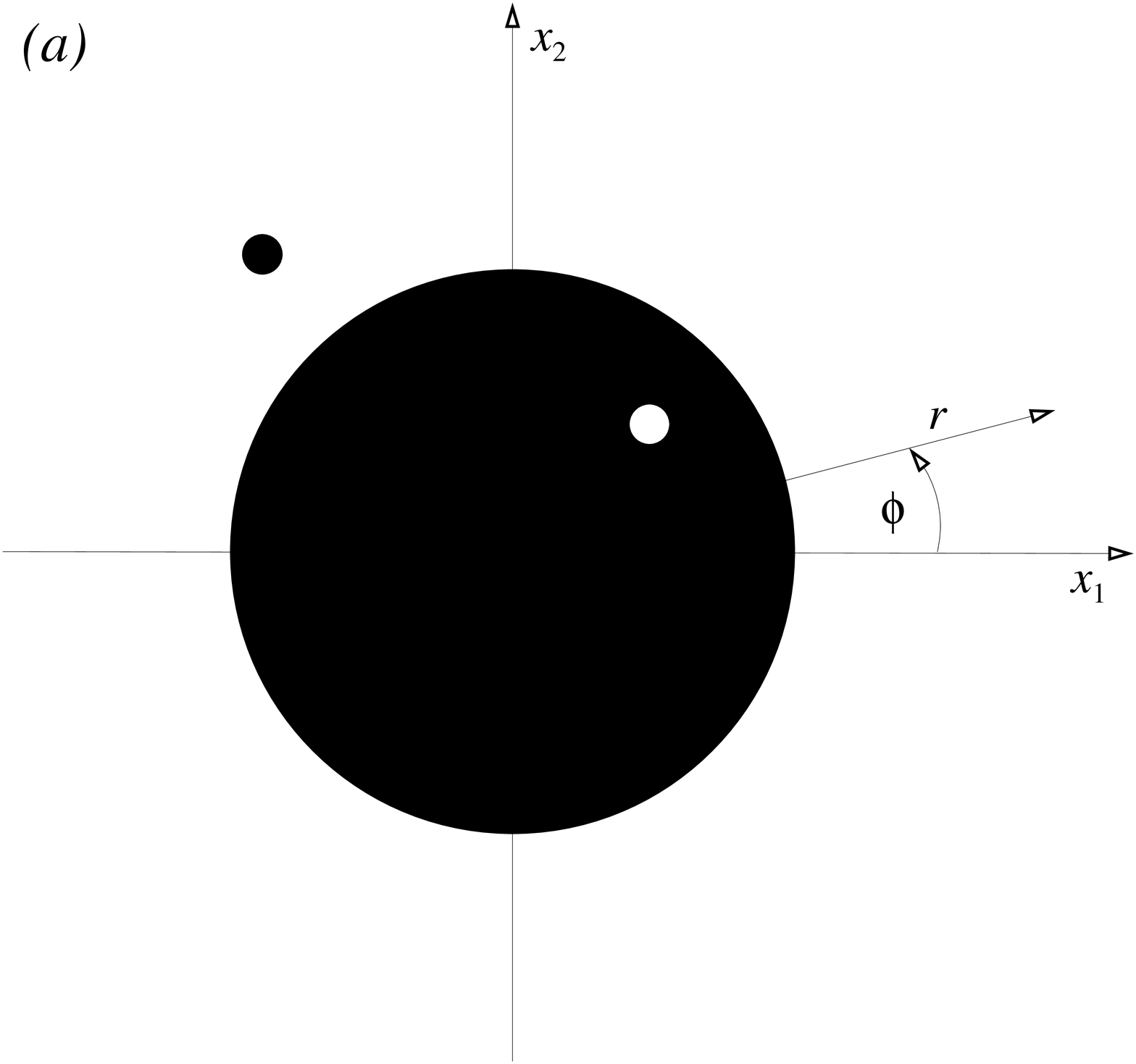}\hspace{1cm}\includegraphics[scale=0.2]{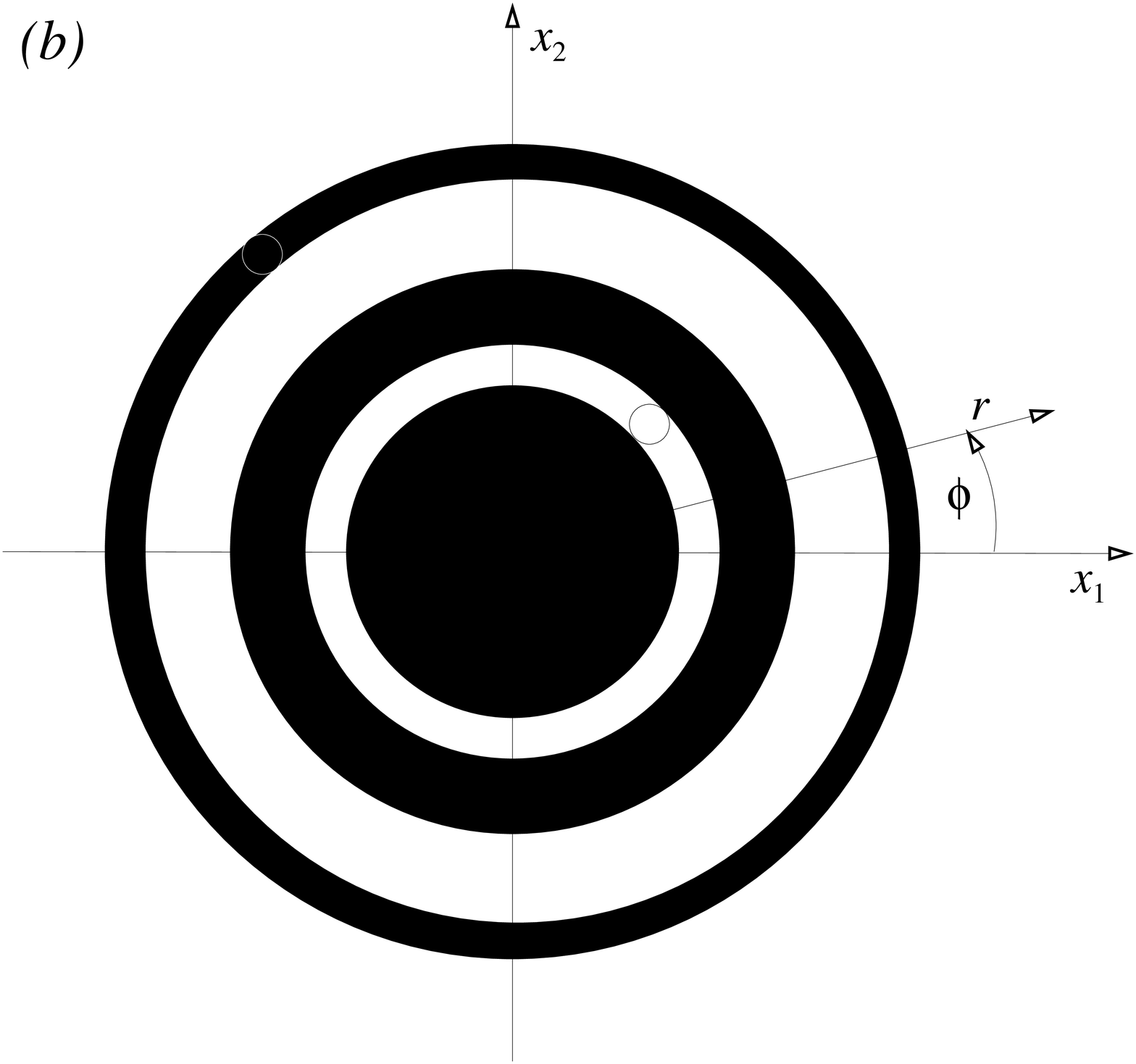}
\end{center}
\caption{\small {\it (a)}: The  configurations which correspond to giant gravitons on the $S^5$ (the hole), and giant gravitons on the AdS$_5$ (the particle). They all have angular momentum in the $\tilde{\phi}$ direction on the $S^5$. {\it (b)}: Smearing the giant (and dual giant) gravitons of part {\it (a)}.}
\label{fig:holey}
\end{figure}

A small hole inside the disc corresponds to a giant
graviton\cite{McGreevy:2000cw}, {\it i.e.,} some D3--branes wrapped on
an $S^3$ within the $S^5$. See figure~\ref{fig:holey}{\it (a)}. It is a
spinning configuration, moving with constant angular speed on the
polar coordinate $\phi$ of the $S^5$.  Notice that how much the hole
is radially displaced in the plane sets where in $\theta$ the giant
graviton is located, and the size of the ${\tilde S}^3$ changes with
radius. Radius translates into angular momentum.  Notice that the
limit to the size of the giant graviton (since it must fit inside the
$S^5$)\cite{McGreevy:2000cw}) is very easy to visualize in this
picture: The hole can only live inside the disc.

Placing a small droplet at some place outside of the disc (see
figure~\ref{fig:holey}{\it (a)}) corresponds to a dual giant
graviton\cite{Grisaru:2000zn,Hashimoto:2000zp}, {\it i.e.,} some
D3--branes wrapping an $S^3$ of AdS$_{5}$. At the edge of the disc, it
has zero size (since the $S^3$ is of zero radius there) but then it
can be moved to arbitrary $\rho$: It can be given arbitrary angular
momentum in the $S^5$.

That the droplets represent spinning objects moving in ${\tilde\phi}$
is clear from the change of variables given in equation~\reef{eq:cov},
since they are at fixed ${\phi}$. Any solution which is symmetric in
$\phi$ will be time--independent. The disc discussed above is the
simplest example. For another example, we can construct a solution
representing a distribution of dual (AdS$_5$) giants by smearing a
droplet into a ring. For a thin ring, this is a good description.
Amusingly, for an increasingly thicker ring, it is increasingly valid
to think of this configuration as representing smeared $S^5$ giants in
the context of a new AdS$_5\times S^5$ geometry whose scale is set by
the size of the outer radius of the ring. Clearly, similar statements
can be made about the case of smearing giants into a thick ring.  See
figure~\ref{fig:holey}{\it (b)}).

\subsection{Motivation}

These coordinates are extremely powerful for describing 1/2--BPS
states and deserve to be explored and better understood. This is one
of the goals of this project. Generally, there are very simple
questions in the dual gauge theory which at times have seemed rather
hard to capture in supergravity. Many pieces of work, often guided by
natural probes of the geometry and the gauge theory (see {\it e.g.,}
refs.\cite{Buchel:2000cn,Evans:2000ct,Johnson:2001ze,Johnson:2000ic})
have shown that this often boils down to a matter of finding good
coordinates on the gravity side. Recent work has shown that many
seemingly complicated gravity duals of {\it e.g.}  various holographic
RG flows\cite{Pilch:2003jg,Pilch:2004yg,Halmagyi:2004jy}, and now
these 1/2--BPS state geometries\cite{Lin:2004nb}, are in fact very
simple, if expressed in the right coordinates. The solutions are
seeded by a simple function, with non--trivial dependence on a
submanifold of the ten--dimensional geometry, satisfying a
differential equation (here it is equation~\reef{eq:diffy}).  On the
basis of probe computations, it has been
suggested\cite{Carlisle:2003nd} that these sorts of coordinates are
probably pointing to an underlying reduced model (perhaps a matrix
model or integrable system) and this is precisely what the coordinates
of ref.\cite{Lin:2004nb} have shown, for the case of 1/2--BPS states,
by making the gauged harmonic oscillator model of
ref.\cite{Berenstein:2004kk} extraordinarily manifest in the geometry.

Whether such a direct uncovering within the supergravity geometry of
such a key ingredient of the reduced model (such as a phase space of
the fermion basis of the matrix model) can be achieved for geometries
which preserve fewer supersymmetries is an issue deserving further
investigation.  This is one the main motivations for studying the
geometry of ref.\cite{Lin:2004nb} further in this paper.

It is families of concentric rings which we will consider in this
paper, using them as a background in which we place spinning strings.
These strings are known\cite{Gubser:2002tv} to be dual to certain near
BPS operators in the gauge theory, about which much has been learned
in recent times\footnote{For a review and thorough discussion of this
  extensive subject, see ref.\cite{Tseytlin:2004xa}.}.  We will find
in this paper that the backgrounds we prepare allow these strings (and
hence operators in the dual theory) to take on new types of behaviour.
For example, in addition to spin and R--charge, they can have orbital
angular momentum as well.  Further, their characteristic dependence
for large and small spin (established in ref.\cite{Gubser:2002tv}) is
joined by an intermediate regime where they behave differently. There,
depending upon the details of the background, they can split and join,
redistributing the angular momentum and R--charge in various ways.
They also have a high energy/spin regime which has a different scaling
behaviour than that known for the long string regime, which exhibits
logarithmic scaling with spin for the anomalous dimension of the
associated operator\cite{Gubser:2002tv}. In the next few sections, we
show how to uncover these new pieces of physics.

\section{Smeared Giants and Concentric Rings}
First, let us show how to describe families of concentric rings. This
is a review of material already presented in ref.\cite{Lin:2004nb}.
The metric with polar coordinates in the $(x^1,x^2)$ plane is:
\begin{equation}
 ds^2=-h^{-2}(dt+V_{r}dr+V_{\phi}d\phi)^{2}+
h^{2}(dy^2+dr^{2}+r^{2}d\phi^{2})+ye^{G}d\Omega_{3}^2+ye^{-G}d\widetilde{\Omega}_{3}^2\ ,
\label{eq:metric}
\end{equation}
If we define $\widetilde{z}=z-\frac{1}{2}$, it can be
shown\cite{Lin:2004nb} that the solutions for $\widetilde{z}$, $V_{r}$
and $V_{\phi}$ are:
\begin{eqnarray}
\widetilde{z}&=&\sum_{i}(-1)^{i+1}\left(\frac{r^{2}-(r^{(i)}_{0})^{2}+y^{2}}{2\sqrt{(r^{2}+(r^{(i)}_{0})^{2}+y^{2})^{2}-4r^{2}(r^{(i)}_{0})^{2}}}-\frac{1}{2}
\right),\nonumber\\
V_{r}&=&0,\nonumber \\
V_{\phi}&=&\frac{1}{2}\sum_{i}(-1)^{i}
\left(\frac{r^2+y^2+(r^{(i)}_{0})^{2}}{\sqrt{(r^{2}+(r^{(i)}_{0})^{2}+y^{2})^{2}-4r^{2}(r^{(i)}_{0})^{2}}}-1\right)
\labell{eq:therings}
\end{eqnarray}
Here $r^{(1)}_0$ is the radius of the outermost circle, $r^{(2)}_{0}$
the next one and so on. In the case of one radius, $r_0$, this is just
the case of AdS$_{5}\times S^5$.

\section{Spinning Strings}
\label{sec:spinningtrings}

Our next step is to study configurations of spinning strings in the
geometry. The configurations which we study always the strings at have
$r=0$. This symmetric situation will allow for a dramatic
simplification of the equations governing the configurations. Having
$r=0$ will mean that we keep the string at $\theta=\pi/2$ so that we
can have non--trivial $\rho$ dependence.

If we parameterize the $S^{3}$ and $\widetilde{S}^{3}$ with the
standard Euler angles: ($\alpha, \beta, \gamma$), and
($\widetilde{\alpha}, \widetilde{\beta}$, $\widetilde{\gamma}$)
respectively, then it can be shown that the following ansatz is
compatible with the equations of motion derived from the sigma model
action:
\begin{eqnarray}
t&=&\kappa\tau \ ;\quad \phi=0\ ;\quad \alpha=\widetilde{\alpha}=0\ ; \nonumber\\
r&=&\dot{y}=0\ ; \quad \beta+\gamma=\omega\tau\ ;\quad \widetilde{\beta}+\widetilde{\gamma}=\widetilde{\omega}\tau\ .
\end{eqnarray}
We denote the world--sheet time and space coordinates by $\tau$ and
$\sigma$ respectively.

For this ansatz the only non--trivial equation of motion is the one
for $y(\sigma)$ and it has the form:
\begin{gather}
 \frac{y''}{y\cosh{G}}+\partial_y\left(\frac{1}{2y\cosh{G}}\right)y'^{2}=\partial_{y}(2y\cosh{G})\kappa^{2}
 -\partial_{y}(ye^G)\omega^2-\partial_{y}(ye^{-G})\widetilde{\omega}^{2}\ .
\end{gather}
However it is more convenient to work with the Virasoro constraint
which in this case is the first integral of the equation of motion
and can be written as:
\begin{gather}
\frac{y'^{2}}{e^{G}+e^{-G}}+y^{2}\left(e^{G}\omega^{2}+e^{-G}\widetilde{\omega}^{2}-\kappa^{2}(e^{G}+e^{-G})\right)=0\ .
\end{gather}
Now using the fact that
\begin{equation}
e^{G}=\sqrt{\frac{1+\widetilde{z}}{-\widetilde{z}}}\ ,
\labell{eq:geezee}
\end{equation}
and after a bit of algebra we arrive at the equation:
\begin{gather}
y'^{2}=y^{2}\left(\frac{a}{\widetilde{z}}+\frac{b}{1+\widetilde{z}},\right)\ ,
\labell{eq:primary}
\end{gather}
where:
\begin{eqnarray}
\widetilde{z}=\sum_{i}(-1)^{i}\frac{{(r^{(i)}_{0})^{2}}}{{(r^{(i)}_{0})^{2}}+{y^{2}}}\ ,\quad\mbox{\rm and}\quad 
a=\omega^{2}-\kappa^2\ ,\quad b=\kappa^2-\widetilde{\omega}^2\ .
\labell{eq:define}
\end{eqnarray}
If we consider the case $a>0, b>0$ (and we can do so without loss of
generality) then we can write:
\begin{equation}
y'^{2}=\frac{y^{2}a}{1+{\widetilde z}}\left(1+\eta+\frac{1}{\widetilde{z}}\right)\ ,\qquad \eta=\frac{b}{a}\ .
\labell{eq:main}
\end{equation}
Note that $-1<\widetilde{z}<0$. 

\begin{figure}[ht]
\begin{center}
\includegraphics[scale=0.5]{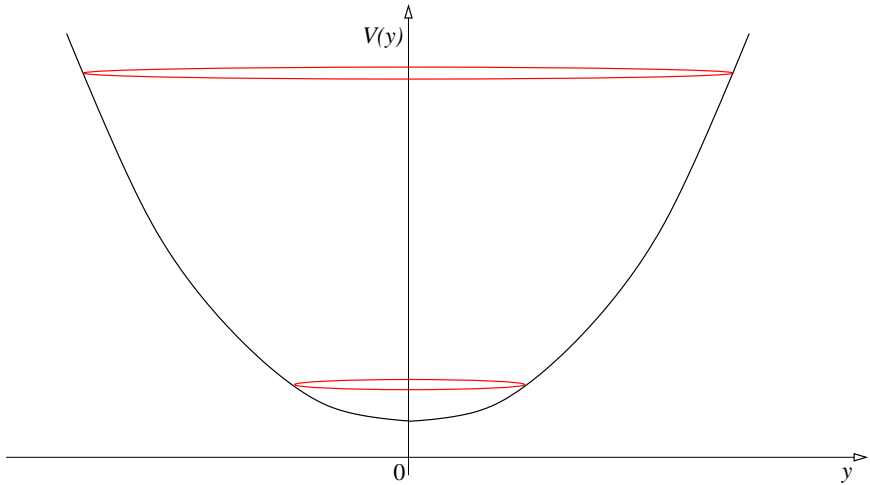}
\end{center}
\caption{\small The effective potential for spinning string configurations in AdS$_5\times S^5$. Shown is a short string and a long string.}
\label{fig:simplepotential}
\end{figure}

Now the condition for having a closed (folded) string is equivalent to
the condition for having a bound state of the effective one
dimensional motion described by~\reef{eq:main}.  Since the factor
outside the brackets is always positive, and since $1/{\widetilde z}$
is negative, we have a natural effective potential problem where the
``energy'' is set by the parameter ${\cal E}=1+\eta$ and the effective
potential is:
\begin{equation}
V(y)=-\frac{1}{\widetilde z}\ .
  \labell{eq:potential}
\end{equation}
The well--known case\cite{Gubser:2002tv} of a folded string in
AdS$_5\times S^5$ is obtained by setting all the $r_0^{(i)}$ to zero
except for one, called $r_0$.  The effective potential is then simply
$1+y^2/r_0^2$. The physics is then straightforward. The folded string
configuration corresponds to a solution $y(\sigma)$ representing a
string of finite extent stretching between the extremes given by the
turning points of our associated particle problem. See
figure~\ref{fig:simplepotential}.

The turning points are further apart (the folded string is longer) as
the parameter $1+\eta$ is increased.  In the limit where the string is
short compared to the scale set by $r_0$ and the setting is
effectively in flat space. The characteristic Regge behaviour, where
the energy of the string scales as the square root of the spin, is
recovered in this limit. There is an opposite limit where the strings
are long compared to the scale set by $r_0$. There, the AdS physics is
important, and the resulting string's energy scales linearly with the
spin, with logarithmic subleading behaviour\cite{Gubser:2002tv}. We
will recover all of this explicitly shortly.

The key point for us is that the new coordinates allow us to explore
more complicated background configurations (or background operators in
the dual gauge theory) quite simply. The new effective potentials
obtained allow for {\it multiple turning points} as a function of
${\cal E}$.  This will introduce many new allowed configurations, and
new physics.

\subsection{Sample Backgrounds}
\begin{figure}[ht]
\begin{center}
\includegraphics[scale=0.25]{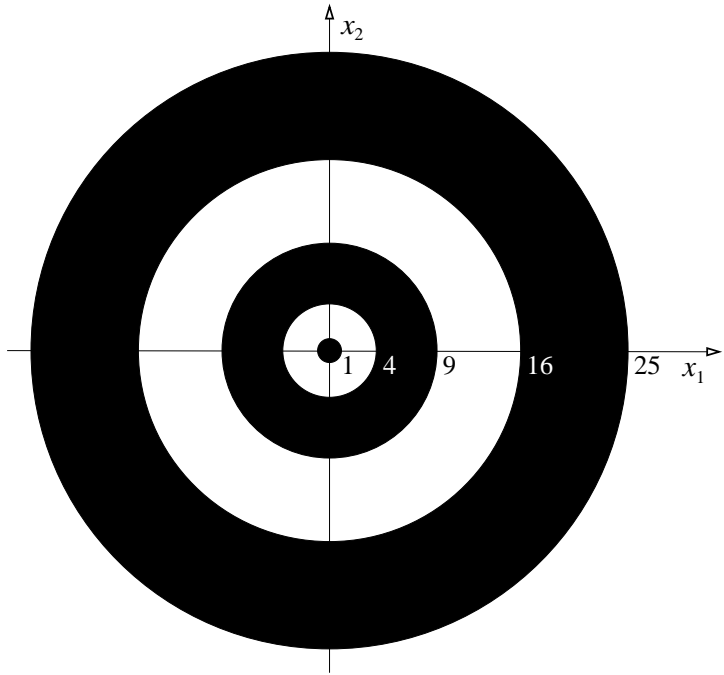}
\end{center}
\caption{\small A set of rings, for Configuration 1, with $m=2$.}
\label{fig:rings}
\end{figure}

\begin{figure}[ht]
\begin{center}
\includegraphics[scale=0.5]{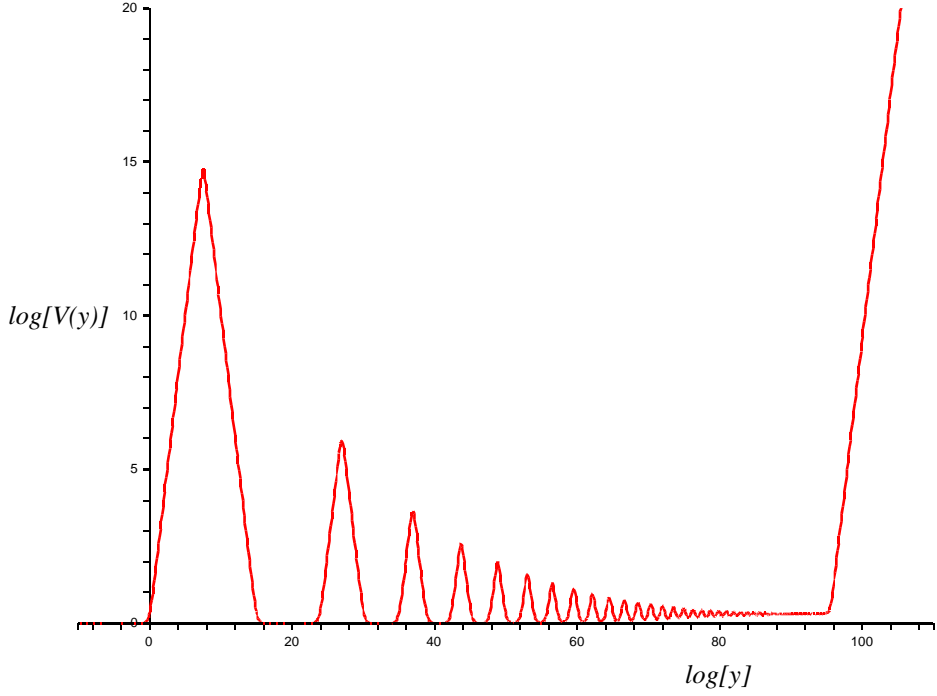}
\end{center}
\caption{\small The effective potential $V(y)$ for the case of Configuration 1, with $m=50$, and $n=81$.}
\label{fig:potentialone}
\end{figure}

\begin{figure}[h]
\begin{center}
\includegraphics[scale=0.5]{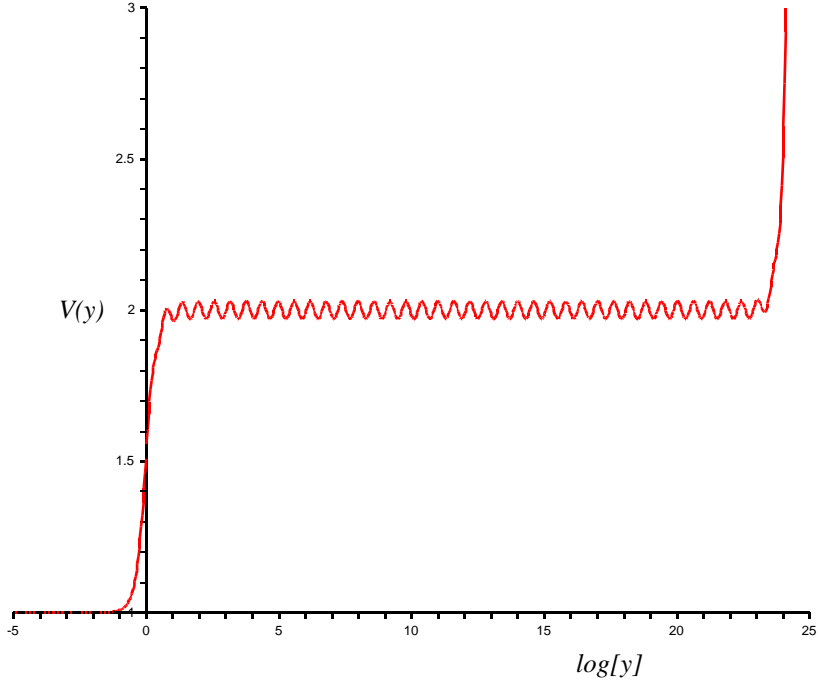}
\end{center}
\caption{\small The effective potential $V(y)$ for the case of Configuration 2, with $q=2$, and $n=81$. }
\label{fig:potentialtwo}
\end{figure}

Two of our favourite configurations are as follows. Set $n$ to be the
number of radii, which we choose to be odd, so as to have a black disc
in the middle of the configuration and $n-1$ rings. The case of one
radius will just be the original AdS$_5\times S^5$.  The
configurations are:
\begin{eqnarray}
\mbox{\rm Configuration 1}:&& r_0^{(i)}=i^m\ , \qquad m\in\IZ^+ \ .\nonumber\\
\mbox{\rm Configuration 2}:&& r_0^{(i)}=q^{(i-1)}\ ,\qquad q\in \IR\ .
  \labell{eq:configurations}
\end{eqnarray}
A four--ring example of Configuration 1 is drawn in
figure~\ref{fig:rings}.  Samples of the effective potentials $V(y)$
resulting from these configurations, for a large number of rings, have
been plotted in figures~\ref{fig:potentialone}
and~\ref{fig:potentialtwo}. (Beware the use of logarithmic scales on
some axes.)  Notice the existence of many minima, in addition to the
overall minimum at $y=0$. These introduce new physics.  Notice also
that for large enough $y$ we recover the monotonic rise associated to
the large $\rho$ (or $y$) behaviour of the overall AdS geometry.

Among the new pieces of physics are the following: First, there are
solutions (pairs of turning points) corresponding to folded strings
with centres of mass which are located away from $y=0$. For these
solutions the angular momentum splits into a spin, $S$, about the
centre of mass, and an orbital angular momentum piece, $L$, since
their motion on the $S^3$ is at finite $y$. This orbital angular
momentum survives even in the point particle limit, in contrast to the
case of the centred string. In the case of the latter, the point
particle limit gives a BPS state in the limit, since $S$ and $L$
vanish. For the non--centred strings, their finite orbital angular
momentum takes them off BPS. Note that for Configuration 1 (above),
the potential is such that the deviation from BPS for the point
particle configurations found in its extra minima is tunably small,
since the minima are close to degenerate with the $y=0$ minimum. The
``energy'' in the potential is set by $\eta$, which can be tuned
arbitrarily close to zero. Configuration 1 allows us to describe a
family of minima which are almost degenerate with the global minimum.
We will discuss this further later.

\subsection{Short Strings and Regge Trajectories}
\begin{figure}[htb]
\begin{center}
\includegraphics[scale=1.0]{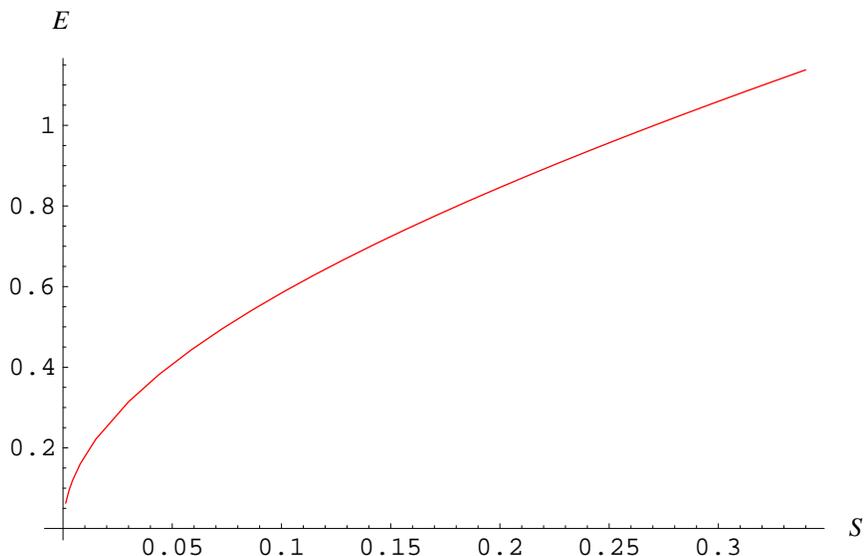}
\end{center}
\caption{\small Plots of $E$ {\it vs} $S$  showing Regge behaviour for a spinning string in the short string regime.  }
\label{fig:Regge}
\end{figure}

Let us consider the case of a short string bound in the region between
$y=0$ and the first zero of the term in the brackets in
equation~(\ref{eq:main}), located at $y=y_0$. Alternatively, we can
think of this bound state as lying between $-y_0$ and $y_0$. See the
lowest central string in figure~\ref{fig:splitting}. Our expressions
for the energy, $E$ and the angular momentum $K$ of the string are:
\begin{align}
  E=\frac{\kappa}{2\pi\alpha'}\int\limits^{2\pi}_0d\sigma{
    \left(2y\cosh{G}\right)}\ ,\qquad
  K=\frac{\omega}{2\pi\alpha'}\int\limits^{2\pi}_0 d\sigma{\left(ye^{G}\right)}\ .
  \labell{eq:next}
\end{align}
Note that since we are considering an odd number of radii, this means
that the conserved quantity associated with rotation in $S^{3}$ is the
angular momentum of the particle (this is so because we are at center
of the ring system $r=0$). Let us expand the left hand side of
equation~\reef{eq:main} to second order in $y$:
\begin{eqnarray}
y'^2&=&\frac{b}{f_{0}}-\left(a-b\frac{f_{1}}{f_{0}^2}\right)y^2+O(y^4)\ ,\nonumber\\
\frac{1}{f_{0}}&=&\sum_{i}(-1)^{i+1}\frac{1}{{r^{(i)}_{0}}^2}\ ,\quad 
\frac{1}{f_{1}}=\sum_{i}(-1)^{i+1}\frac{1}{{r^{(i)}_{0}}^4}\ .
\labell{eq:expanded}
\end{eqnarray}
Therefore the conditions for having a singly folded string with
radius $y_{0}\ll 1$ are:
\begin{align}
a-b\frac{f_{1}}{f_{0}^2}=1&&y_{0}^2=\frac{b}{f_{0}}
\end{align}
Now if we consider rotation only in $S^3$, {\it i.e.,}
$\widetilde{\omega}=0$, then from the definitions of $a$ and $b$ in
(\ref{eq:define}) it follows that:
\begin{align}
\labell{14} y_{0}^2=\frac{\kappa^2}{f_{0}}\ ,\quad
\omega^2=1+\kappa^2\left(1+\frac{f_{1}}{f_{0}^2}\right)\ ,\quad
\kappa\ll1\ .
\end{align}
Expanding the integrands in equation~(\ref{eq:next}) to second order
in $y$ leads to the expressions for the energy, $E$, and spin, $S$:
\begin{align}
\labell{15} E=\frac{\kappa}{2\pi\alpha'}\int\limits^{2\pi}_0
d\sigma{\frac{1}{\sqrt{f_{0}}}}+O(\kappa^3)=\frac{\kappa}{\alpha'}\frac{1}{\sqrt{f_{0}}}+O(\kappa^3)\ ,\nonumber \\
S=\frac{\omega}{2\pi\alpha'}\int\limits^{2\pi}_0
d\sigma{\frac{\kappa^2}{\sqrt{f_{0}}}}\sin^{2}\sigma=\frac{\kappa^2}{4\pi\alpha'\sqrt{f_{0}}}+O(\kappa^4)\ .
\end{align}
From this we see that for short strings ($\kappa\ll1$) we have the
Regge trajectories:
\begin{gather}
\labell{16} E^2=\frac{4\pi\sqrt{f_{0}}}{\alpha'}S\ .
\end{gather}
The quantity $f_0$ given in equation~\reef{eq:expanded}, acts as an
effective AdS radius in the problem in this limit. For the one radius
({\it i.e.,} disc) problem the result of ref.\cite{Gubser:2002tv} for
AdS is reproduced. See figure~\ref{fig:Regge} for a numerical plot of
this behaviour for a three--ring example.

It is possible to tune the location of the ring radii to obtain other
minima, as is evident from our examples in the previous subsection.

\subsection{Particles and Short Strings  at $y\neq 0$}
Consider the solutions corresponding to the extra minima that can be
generated by our effective potential $V(y)=-1/{\widetilde{z}}$.
Denote the position of the $i$th extra minimum by $y_{0}^{(i)}$.

Let us first study the point--like solutions corresponding to string
positioned at the bottom of the potential well, collapsed to a
particle. We can do that by tuning $\eta$ appropriately. Let's denote
its corresponding value by $\eta^{(i)}$.  Then we have the relation
for our point--like configurations:
\begin{gather}
\labell{25}
1+\eta^{(i)}=-\frac{1}{\widetilde{z}(y_{0}^{(i)})}\ .
\end{gather}
Let's compute the conserved quantities related to $\omega,
\widetilde{\omega}$ and $\kappa$. For point--like strings they are
simply (using equation~\reef{eq:geezee}):
\begin{align}
\labell{26}
&E_{0}^{(i)}=\frac{\kappa^{(i)}}{\alpha'}2\cosh{G}|_{y_{0}^{(i)}}=\frac{\kappa^{(i)}}{\alpha'}y_{0}^{(i)}\left(\sqrt{\eta^{(i)}}+\frac{1}{\sqrt{\eta^{(i)}}}\right)\\
&L_{0}^{(i)}=\frac{\omega^{(i)}}{\alpha'}y e^{G}|_{y_{0}^{(i)}}=\frac{\omega^{(i)}}{\alpha'}y_{0}^{(i)}\sqrt{\eta^{(i)}}\\
&J_{0}^{(i)}=\frac{\widetilde{\omega}^{(i)}}{\alpha'}ye^{-G}|_{y_{0}^{(i)}}=\frac{\widetilde{\omega}^{(i)}}{\alpha'}y_{0}^{(i)}\frac{1}{\sqrt{\eta^{(i)}}}
\end{align}
Let us define, for convenience, a parameter
$\zeta=(\frac{\omega}{\widetilde{\omega}})^2$.  It can be verified
that from $\eta\geq 0$ follows $\zeta\geq 1$. We can now check the BPS
condition. After a bit of algebra we find:
\begin{gather}
\labell{27}
\frac{E^{(i)}-J^{(i)}}{J^{(i)}}=\frac{\Delta^{(i)}-J^{(i)}}{J^{(i)}}=\sqrt{1+\eta^{(i)}}\sqrt{1+\zeta^{(i)}\eta^{(i)}}-1\geq0\ .
\end{gather}
Notice that since $\zeta$ is not bounded from above, we can go as far
as we wish from the BPS condition. This is not surprising because now
we are free to increase the (orbital) angular momentum, $L_{0}^{(i)}$,
of the particle in the $S^3$ without changing the corresponding one in
$\widetilde{S}^3$ which is the R--charge $J^{(i)}$. However if we fix
$\zeta$ and tune $\eta^{(i)}$ close to zero, we can make the
corresponding operator as close to BPS as wish. It is natural because
from (\ref{26}) we can see that $L_{0}\rightarrow 0$ as
$\sqrt{\eta}\rightarrow 0$. So in the Configuration 1 example that we
gave, since the minima are all degenerate with the $y=0$ minimum, this
is the case.

We now study  short strings corresponding to the harmonic
approximation for the effective potential around the minima
$y_{0}^{(i)}$. For this purpose we can expand:
\begin{gather}
\labell{28} y=y_{0}^{(i)}+\xi\ .
\end{gather}
Now if we expand our effective potential we have:
\begin{gather}
\labell{29}
-\frac{1}{\widetilde{z}(\xi)}=1+\eta^{(i)}+\frac{\widetilde{z}''(y_{0}^{(i)})}{2\widetilde{z}(y_{0}^{(i)})^2}\xi^2+O(\xi^3)\ .
\end{gather}
Therefore the equation of motion for $\xi$ is:
\begin{eqnarray}
\xi'^2&=&\left.\frac{ay^2\widetilde{z}''}{2(1+\widetilde{z})\widetilde{z}^2}\right|_{y_{0}^{(i)}}(\xi_{0}^2-\xi^2)+O(\xi_{0}^3)\ ,\nonumber\\
 \xi_{0}^2&=&(\eta-\eta_{0}^{(i)})\left.\frac{2\widetilde{z}^2}{\widetilde{z}''}\right|_{y_{0}^{(i)} }
\labell{30}
\end{eqnarray}
Now the periodicity condition (for having singly folded string)
implies:
\begin{gather}
\labell{31}
a=a^{(i)}=\left.\frac{2(1+\widetilde{z})\widetilde{z}^2}{y^2\widetilde{z}''}\right|_{y_{0}^{(i)}}\ ,
\end{gather}
and we have the solution:
\begin{gather}
 y=y_{0}^{(i)}+\xi_{0}\sin\sigma\ .
\labell{eq:solution}
\end{gather}
The next step is to substitute (\ref{eq:solution}) into the general
expressions for the conserved quantities $E$, $K$ and $J$. Before
proceeding, let us note that from here on, in order to simplify the
notation we'll omit the upper index $(i)$ denoting that everything
that we write holds for the $i^{th}$ minimum. It should be clear from
the context.

Only even powers of $\xi$ will contribute under the integral and so
using:
\begin{eqnarray}
e^{G}&=&\sqrt{\eta_{0}}(1+\Omega^2\xi_{0}^2\sin^2{\sigma})+O(\xi_{0}^3)\nonumber\\
e^{-G}&=&\frac{1}{\sqrt{\eta_{0}}}(1-\Omega^2\xi_{0}^2\sin^2{\sigma})+O(\xi_{0}^3)\ ,
\labell{33}
\end{eqnarray}
where we have introduced
\begin{equation}
\Omega^2=\left.\frac{\widetilde{z}''}{4\eta_{0}\widetilde{z}^2}\right|_{y_{0}}\ ,
\end{equation}
we arrive at the energy, total angular momentum, and R--charge:
\begin{align}
&E=\frac{{\kappa}y_{0}}{\alpha'}(\sqrt{\eta_{0}}+\frac{1}{\sqrt{\eta_{0}}})\left(1+\frac{1}{4\pi}\frac{\eta_{0}-1}{\eta_{0}+1}\Omega^2\xi_{0}^2\right)+O(\xi_{0}^4)\ ,\nonumber\\
&K=\frac{{\omega}y_{0}}{\alpha'}\sqrt{\eta_{0}}\left(1+\frac{1}{4\pi}\Omega^2\xi_{0}^2\right)+O(\xi_{0}^4)\ ,\nonumber\\
&J=\frac{{\widetilde{\omega}}y_{0}}{\alpha'}\frac{1}{\sqrt{\eta_{0}}}\left(1-\frac{1}{4\pi}\Omega^2\xi_{0}^2\right)+O(\xi_{0}^4)\ .
\end{align}
Now let's study the case when we don't have R--charge, {\it i.e.,}
$\widetilde{\omega}=0$. Then from equation~(\ref{31}) and the definition for
$\eta$ we get that:
\begin{eqnarray}
\kappa^2&=&a^{(i)}\eta_{0}(1+2\Omega^2\xi_{0}^2)\ ,\nonumber\\
\omega^2&=&a^{(i)}(1+\eta)\left(1+2\frac{\eta_{0}}{1+\eta_{0}}\Omega^2\xi_{0}^2\right)\ .
\end{eqnarray}
Therefore:
\begin{align}
&E=y_{0}\frac{\sqrt{a^{(i)}\eta_{0}}}{\alpha'}(\sqrt{\eta_{0}}+\frac{1}{\sqrt{\eta_{0}}})\left(1+\left(1+\frac{1}{4\pi}\frac{\eta_{0}-1}{\eta_{0}+1}\right)\Omega^2\xi_{0}^2\right)+O(\xi_{0}^4)\ ,\nonumber\\
&K=y_{0}\frac{\sqrt{a^{(i)}(\eta_{0}+1)}}{\alpha'}\sqrt{\eta_{0}}\left(1+\left(\frac{\eta_{0}}{\eta_{0}+1}+\frac{1}{4\pi}\right)\Omega^2\xi_{0}^2\right)+O(\xi_{0}^4)\ .
\labell{eq:expressions}
\end{align}
Now it seems prudent to introduce the notation:
\begin{align}
  &E_{0}=y_{0}\frac{\sqrt{a^{(i)}\eta_{0}}}{\alpha'}
  \left(\sqrt{\eta_{0}}+\frac{1}{\sqrt{\eta_{0}}}\right)\ ,\nonumber\\
  &L_{0}=y_{0}\frac{\sqrt{a^{(i)}(\eta_{0}+1)}}{\alpha'}\sqrt{\eta_{0}}\ ,\nonumber\\
  &S=y_{0}\frac{\sqrt{a^{(i)}(\eta_{0}+1)}}{\alpha'}
  \sqrt{\eta_{0}}\left(\frac{\eta_{0}}{\eta_{0}+1}+\frac{1}{4\pi}\right)\Omega^2\xi_{0}^2\ .
\labell{eq:params}
\end{align}
The quantities $E_{0}$ and $L_{0}$ have the interpretation as the
energy and angular momentum of the string for the point--like limit.
$L_0$ is an orbital angular momentum.  Interestingly, consider the
last quantity in equation~\reef{eq:params} which we have called $S$,
which arises due to the finite size of the string.  It seems
consistent to regard it as the spin of the string, together with a
term which might have an interpretation as a spin--orbit interaction.

Finally, if we take the square of the first quantity in
equation~(\ref{eq:expressions}) and substitute our definitions, we
arrive at the expression:
\begin{gather}
  \labell{38} E^2=E_{0}^2+\frac{4\pi{R_{\rm eff}^2}}{\alpha'}S\ ,
\end{gather}
where we have defined $R_{\rm eff}$ as:
\begin{gather}
\labell{39}
R_{\rm eff}^2=\frac{y_{0}}{2\pi}\sqrt{a^{(i)}\frac{(1+\eta_{0})^3}{\eta_{0}}}\left\{\frac{\eta_{0}+\frac{4\pi-1}{4\pi+1}}{\eta_{0}+\frac{1}{4\pi+1}}\right\}\ .
\end{gather}

\subsection{Long Strings}

Let us consider the very long string limit, where the centred string
has sufficiently high $\eta$ so that it is far above all maxima and so
stretches out to the far reaches of AdS. See the uppermost string in
figure~\ref{fig:splitting} for example.

We have the following expansions in terms of $1/y$:
\begin{eqnarray}
-\frac{1}{\widetilde{z}}&=&\frac{y^2}{g_{0}}+\frac{g_{1}}{g_{0}^2}+O\left(\frac{1}{y^2}\right)\ ,
\quad \frac{1}{1+\widetilde{z}}=1+\frac{g_{0}}{y^2}\ ,\nonumber\\
g_{0}&=&\sum_{i}(-1)^{i+1}{r^{(i)}}^{2}\ ,
\quad g_{1}=\sum_{i}(-1)^{i+1}{r^{(i)}}^{4}\ .
\labell{17}
\end{eqnarray}
This leads to the effective equation of motion for $y$:
\begin{align}
\labell{18} y'^2=\frac{a}{g_{0}}y^2(y_{0}^2-y^2)\ ,\quad
\frac{y_{0}^2}{g_{0}}=\eta-\frac{g_{1}}{g_{0}^2}\ .
\end{align}
Since we are considering long string ($y_{0}\gg\sqrt{g_{0}}$) we
can use $y_{0}^2\approx g_{0}\eta$. In order to proceed we need
expansions for $y{e}^{G}$ and $y{e}^{-G}$:
\begin{align}
  \labell{19}
  y{e}^{G}=\frac{y^2}{\sqrt{g_{0}}}-\frac{g_{0}^2-g_{1}}{2g_{0}^{3/2}}+O\left(\frac{1}{y^2}\right)\ 
  ,\quad y{e}^{-G}=\sqrt{g_{0}}+O\left(\frac{1}{y^2}\right)\ .
\end{align}
Using this we arrive at the following expressions for $S$ and $E$:
\begin{align}
\labell{20}
E=\frac{4\sqrt{\eta}}{2\pi\alpha'}\left(\int\limits^{y_{0}}_{\epsilon}
dy{\frac{y}{\sqrt{y_{0}^{2}-y^{2}}}}+\frac{g_{0}^2+g_{1}}{2g_{0}}\int\limits^{y_{0}}_{\epsilon}
{\frac{dy}{y\sqrt{y_{0}^{2}-y^{2}}}}\right)\ ,\nonumber\\
S=\frac{4\sqrt{\eta+1}}{2\pi\alpha'}\left(\int\limits^{y_{0}}_{\epsilon}
dy{\frac{y}{\sqrt{y_{0}^{2}-y^{2}}}}-\frac{g_{0}^2-g_{1}}{2g_{0}}\int\limits^{y_{0}}_{\epsilon}
{\frac{dy}{y\sqrt{y_{0}^{2}-y^{2}}}}\right)\ .
\end{align}
Now using the fact that:
\begin{eqnarray}
&&\int\limits^{y_{0}}_{\epsilon}
dy{\frac{y}{\sqrt{y_{0}^{2}-y^{2}}}}=y_{0}+O(y_{0})=\sqrt{g_{0}\eta}+O\left(\frac{1}{\eta}\right)^0\ ,\nonumber\\
&&\int\limits^{y_{0}}_{\epsilon}{\frac{dy}{y\sqrt{y_{0}^{2}-y^{2}}}}=\frac{1}{2\sqrt{\eta}}\left(\ln{\eta}+O\left(\frac{1}{\eta}\right)^0\right)\ ,
\end{eqnarray}
and that in this limit  $\sqrt{\eta+1}\approx\sqrt{\eta}$, we get the following:
\begin{eqnarray}
E&=&\frac{4\sqrt{g_{0}}}{2\pi\alpha'}\left(\eta+\frac{g_{0}^2+g_{1}}{4g_{0}^2}\ln{\eta}+O\left(\frac{1}{\eta}\right)^0\right)\ ,\nonumber\\
S&=&\frac{4\sqrt{g_{0}}}{2\pi\alpha'}\left(\eta-\frac{g_{0}^2-g_{1}}{4g_{0}^2}\ln{\eta}+O\left(\frac{1}{\eta}\right)^0\right)\ .
\end{eqnarray}
Notice that the energy and spin diverge with the length of the string,
since $y_0^2\propto\eta$. This is a natural state of affairs. In
physical terms, by eliminating $\eta$ we get the leading order
long string behaviour:
\begin{gather}
  E-S=\frac{\sqrt{\tilde\lambda}}{\pi}\ln{S/\sqrt{\tilde\lambda}}+\dots
  \labell{eq:longstring}
\end{gather}
Here we have defined $\sqrt{\tilde\lambda}=\sqrt{g_{0}}/\alpha'$. The
quantity $g_0$ is now the effective radius. Once again, note that the
case of one radius, the disc, returns us to the known AdS result of
ref.\cite{Gubser:2002tv}.

\subsection{Intermediate Behaviour and Splitting Strings}
\begin{figure}[ht]
\begin{center}
\includegraphics[scale=0.5]{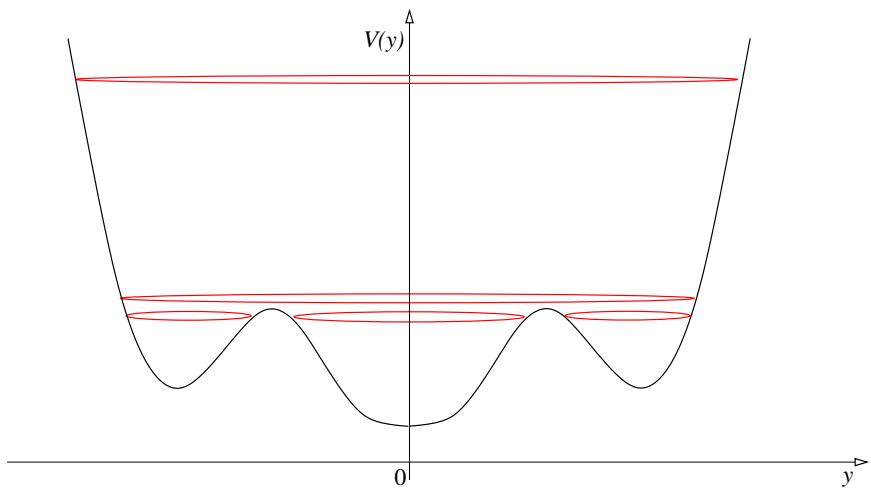}
\end{center}
\caption{\small The effective potential for spinning string configurations resulting from two rings. A long string configuration can split into smaller strings, and smaller strings can merge into longer strings.}
\label{fig:splitting}
\end{figure}

We now turn to more new behaviour, which occurs in the intermediate
regime. The new behaviour can be arrived at as follows.  Imagine
reducing the spin and energy of one of the very long strings from the
previous subsection by reducing $\eta$. Eventually, new behaviour will
occur when a new zero appears in the following quantity:
\begin{equation}
{\cal E}-V(y)=1+\eta+\frac{1}{\widetilde z}\ .
  \labell{eq:eff}
\end{equation}
When this occurs, our string will be degenerate with a {\it trio} of
strings with the same $\eta$. See figure~\ref{fig:splitting}. The
string must split into these strings at this juncture. Notice that
although the two outermost daughter strings are not centered and so
carry orbital angular momentum, they carry equal and opposite amounts,
and so the change in $L$ is zero.  Various local versions of this
behaviour can happen as well, with one non--centred string splitting
into two at some finite $y$. The splitting can occur in such a a way
as to preserve the total angular momentum, by distributing it into
spin an orbital angular momentum as appropriate.

The splitting is characterized by the occurrence of a new zero in the
quantity in expression~\reef{eq:eff}, {\it together with a maximum of
  the effective potential} at that zero. This is new. Its effect is to
give a divergence of the quantities $E$ and $S$, although the length
of the string is not itself diverging. This is contrast to the
divergence encountered in the long string case in the previous section
which occurred for diverging length.  The divergence here is for
finite length, and is the characteristic of the string wishing to
break into smaller strings, or join with others to make a larger one
(depending upon whether one is approaching the maximum from above or
below).

The origin of the divergence can be simply traced to the fact that the
integrals all have the measure:
\begin{equation}
  \label{eq:measure}
  d\sigma=\frac{\sqrt{{\widetilde z}+1}}{y\sqrt{a}}\frac{dy}{\sqrt{{\cal E}-V(y)}}\ ,
\end{equation}
and near the splitting point, $y_s$, we have:
\begin{equation}
  \label{eq:splitpoint}
  V(y)={\cal E}-\frac{|V^{\prime\prime}(y_s)|}{2}(y-y_s)^2+O(y-y_s)^3\ ,
\end{equation}
and since the rest of the integrand in the expressions~\reef{eq:next}
for $E$ and $S$ involve no special features in the neighbourhood of
the splitting point:
\begin{eqnarray}
\label{44}
E&=&\frac{\sqrt{\eta}}{2\pi\alpha'}\int^{y_{0}}dy\,\,
2\cosh{G}\sqrt{\frac{1+\widetilde{z}}
{\left(1+\eta+\frac{1}{\widetilde{z}}\right)}}\ ,\nonumber\\
S&=&\frac{\sqrt{\eta+1}}{2\pi\alpha'}\int^{y_{0}}dy\,\,e^{G}
\sqrt{\frac{1+\widetilde{z}}
{\left(1+\eta+\frac{1}{\widetilde{z}}\right)}}\ ,
\end{eqnarray}
there is {\it always a logarithmic divergence} in these quantities as
$y_0$ (one of the turning points for a configuration) approaches the
splitting point $y_s$ (a turning point which is also a maximum of
$V(y)$).

This results in a new characteristic behaviour for $E$ {\it vs} $S$
for our operators near these points.  They both diverge
logarithmically as the quantity $\epsilon=y_s-y_0$ vanishes:
\begin{eqnarray}
&&E(\epsilon)=C(\eta_{cr}+1)\ln{\frac{1}{\epsilon}}+O(\epsilon^0)\ ,\nonumber\\
&&S(\epsilon)=C\sqrt{(\eta_{cr}+1)\eta_{cr}}\ln{\frac{1}{\epsilon}}+O(\epsilon^0)\ ,\end{eqnarray}
where 
\begin{equation}
C\equiv\frac{1}{\pi\alpha'}\left.\sqrt{\frac{2(1+\widetilde{z})\widetilde{z}^2}{\widetilde{z}''}}\right|_{y_{s}}\ ,
\end{equation}
and so we have a leading {\it linear} relation between $E$ and $S$ in the
split/joint regime:
\begin{gather}
 E=\left(\frac{\eta_{cr}+1}{\eta_{cr}}\right)^{\frac12}S+O(S^0)\ ,
\labell{eq:newscaling}
\end{gather}
where $\eta_{cr}$ is given by:
\begin{gather}
 1+\eta_{cr}=-\frac{1}{\widetilde{z}(y_{s})}\ .
\end{gather}
This is a characteristically different behaviour from the behaviour
observed in the long string regime\cite{Gubser:2002tv}, showing the
logarithmic behaviour of the anomalous dimension of the associated
operators, presented in equation~\reef{eq:longstring}. It would be
interesting to reproduce this behaviour in the dual gauge theory.

\begin{figure}[htb]
\begin{center}
\includegraphics[scale=1.0]{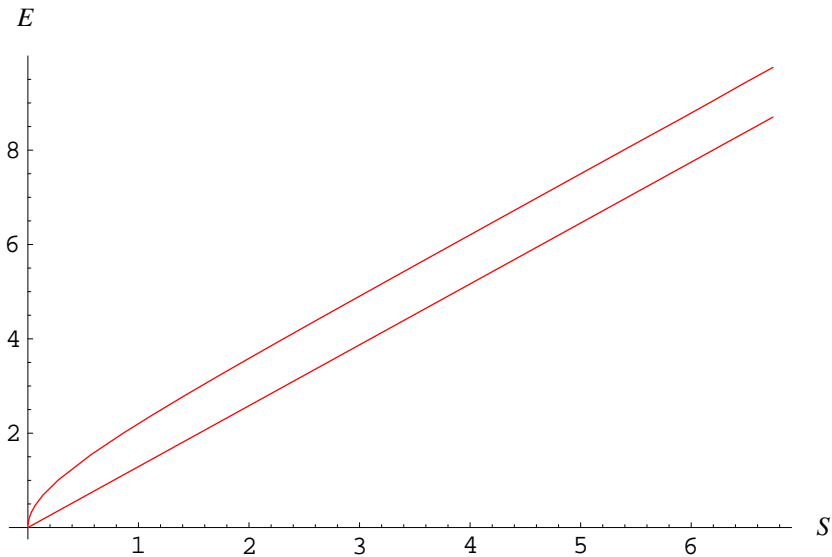}
\end{center}
\caption{\small Plots of $E$ {\it vs} $S$  for a non--centred string including the short string regime (Regge) at low energy and the linear ``splitting regime'' at high energy. The upper curve is the result of numerically evaluating the quantities defined by the integrals in the text. The lower curve is a  plot of the  leading linear term of  the analytic result of equation~\reef{eq:newscaling} for the splitting regime.  }
\label{fig:general}
\end{figure}

We plotted graphs of $E$ {\it vs} $S$ for a non--centred string in
figure~\ref{fig:general}, using a three--ring example, with the radii
tuned so as to produce a minimum in $V(y)$ away from $y=0$. (The
three--ring example we used for this plot had $r_0^{(1)}=1,
r_0^{(2)}=4$, and the third ring at a very much larger distance so as
to cleanly isolate the maximum in $V(y)$.)

\section{Closing Remarks}
The new coordinates of ref.\cite{Lin:2004nb} have made it easy to
study rather interesting configurations of smeared giant gravitons as
spinning string backgrounds. These backgrounds allow the strings to
take on more interesting behaviour than in the case of pure AdS. We
were able to uncover this because we designed an ansatz for the
spinning string configuration which allowed for a reduction to a
simple one--dimensional mechanical system. A wide range of interesting
potentials for this problem can be engineered by adjusting the size
and distributions of the smeared gravitons. In particular, the fact
that the potential can have multiple minima yielded a variety of new
behaviours for the strings. This all translates rather
straightforwardly into statements about the associated operators in
the gauge theory, and it would be interesting to compare to
computations within the field theory with a preparation of the
corresponding background chiral primaries with large R--charge.

The description of the splitting and joining of the strings is
intriguingly simple. In fact, as it reduces to the problem of studying
the coalescence of branch cuts of a simple function, it is very
similar in spirit to the ingredients which arise in solving some large
$N$ matrix models\cite{Brezin:1977sv}. There, a divergence occurs as
the model tries to make the transition from one--cut support to
multiple cuts, just as we saw here for the splitting of the string.
Perhaps this connection holds some useful clues for other
applications. Another interesting question is whether there is any
sense to be made of quantum mechanical effects in this effective
potential. Would tunneling effects between neighbouring minima
correspond to new decay channels in the full type~IIB background and
hence in the dual gauge theory? This is also worth
exploring\footnote{While we were writing this report on our work,
  another paper appeared which studies the splitting of spinning
  strings in AdS\cite{Peeters:2004pt}. We do not know if there is a
  direct connection with our results.}.

Finally, the questions we mentioned in the introduction deserve to be
repeated here. Are there accessible physical manifestations of an underlying
reduced model (analogous to the matrix model's fermionic phase space
for this 1/2--BPS situation)  in appropriate
coordinates for supergravity duals of Yang--Mills theory with fewer
supersymmetries? Might they serve as powerful analytical tools for a
class of interesting physical questions? The intriguing way in which
the coordinates of ref.\cite{Lin:2004nb} operate is cause for
optimism.

\label{sec:discussion}

\section*{Acknowledgments}
VF was supported in part by the U.S. Department of Energy under grant
\# DE--FG03--84ER--40168.  VF and CVJ thank Iosef Bena for leading a
very useful USC theory group meeting about ref.\cite{Lin:2004nb}, and
CVJ thanks both Iosef Bena and David Berenstein for leading an
excellent discussion of ref.\cite{Lin:2004nb} at the ITP workshop on
QCD and String Theory.

\providecommand{\href}[2]{#2}\begingroup\raggedright\endgroup

\end{document}